# UV plasmonic properties of colloidal liquid-metal eutectic gallium-indium alloy nanoparticles

Philipp Reineck[a], Yiliang Lin[b], Brant C. Gibson[a], Michael D. Dickey[b], Andrew D. Greentree[a], and Ivan S. Maksymov[a]

[a] ARC Centre of Excellence for Nanoscale BioPhotonics, School of Science, RMIT University, Melbourne, VIC 3001, Australia
[b] Departments of Chemical and Biomolecular Engineering and Chemistry, North Carolina State University, Raleigh, NC 27695, USA

**Nanoparticles made of non-noble metals such as gallium have recently attracted significant attention due to promising applications in UV plasmonics. To date, experiments have mostly focused on solid and liquid pure gallium particles immobilized on solid substrates. However, for many applications, colloidal liquid-metal nanoparticle solutions are vital. Here, we experimentally demonstrate strong UV plasmonic resonances of eutectic gallium-indium (EGaIn) liquid-metal alloy nanoparticles suspended in ethanol. We rationalise experimental results through a theoretical model based on Mie theory. Our results contribute to the understanding of UV plasmon resonances in colloidal liquid-metal EGaIn nanoparticle suspensions. They will also enable further research into emerging applications of UV plasmonics in biomedical imaging, sensing, stretchable electronics, photoacoustics, and electrochemistry.**

Gold and silver have dominated plasmonics in the past decade and remain highly relevant to the field.[1] However, many other plasmonic materials exist, which have significant advantages in several emerging applications. Such plasmonic materials include aluminium,[2] copper,[3] platinum and palladium,[4] as well as ferromagnetic metals and their alloys such as cobalt, nickel, iron and Permalloy.[5] Aluminium for example, shows strong plasmon resonances from the visible to the ultraviolet (UV) spectral region. It is abundant in nature, used industrially on a large scale and therefore inexpensive.[2] Copper enables ultralow-loss plasmonic waveguides that can outperform their counterparts made of gold and is also compatible with both silicon photonics and silicon microelectronics.[3] Platinum and palladium are essential for plasmonic optical gas sensing.[6] The plasmonic properties of ferromagnetic metals can conveniently be controlled via external magnetic fields, which is the basis of the field of magneto-plasmonics.[5] Furthermore, this allows combining plasmonics with magnonics and spintronics.[7]

The interest in plasmonic metals other than gold and silver is also rooted in their ability to support plasmon resonances in the UV spectral range. Important examples are gallium,[8–16] indium,[9,10] tin,[9,10] thallium,[9,10] antimony[14], lead,[10] bismuth,[14,17] magnesium,[18] and rhodium,[10,19,20] as well as some of their alloys.[21] Nanoparticles (NPs) made of these metals are promising candidates for several specific applications in biomedical imaging and sensing. For example, many biomolecules such as DNA strongly absorb light in the UV spectral range. Plasmon-enhanced UV spectroscopy may provide a means to observe dynamic biochemical processes and characterise the chemical properties of these molecules *in situ*. Moreover, UV-plasmonic NPs have been predicted to find applications in biomedicine as well as in label-free DNA and single molecule sensing[22–25]. In cardiovascular science, plasmonic NPs also play an important role in the imaging of atherosclerotic plaques and the study of their pathophysiology with light at different wavelengths including the UV spectral range.[26–28]

Among UV-plasmonic NPs known today, those made of pure gallium have attracted particular attention[8,9,12,13,16,25,29]. Gallium is an environmentally stable liquid metal at near room temperature and it has been a key element in both electronic and optoelectronic devices since the 1960s.[30] However, despite being well-understood, gallium as well as its alloys continue to be a topic of active research revealing new and unique properties of this material.[31–34] Unlike gold and silver, gallium has a Drude-like dielectric permittivity function extending from the UV range through the visible and, mostly in the liquid state, into the infrared spectral region.[13,21] Due to their high environmental stability and excellent mechanical properties gallium and its alloys are highly relevant for many emerging applications[32–34] including reconfigurable UV-plasmonic devices.[35] The liquid nature of gallium alloys opens unique applications in reconfigurable and photoacoustic UV-plasmonics devices.[35]

The field of gallium-based UV plasmonics has thus far mostly focused on solid and liquid NPs fabricated on solid substrates. However, since the melting point of gallium is well below the normal human body temperature, understanding and engineering *colloidal* liquid gallium NPs is vital for biomedical applications.[22–25]

In this paper, we experimentally and theoretically investigate eutectic gallium-indium (EGaIn, 75% Ga 25% In by weight, ~15.5°C melting point[29]) spherical NPs suspended in ethanol and demonstrate their strong UV plasmonic resonances. Our numerical model based on Mie theory also predicts UV plasmonic resonances and is in good qualitative agreement with experimental results. We also suggest that for some UV plasmonic applications it would be more advantageous to utilize EGaIn than pure gallium because of its superior mechanical, electrical and chemical properties as well as ease of nanofabrication. Indeed, EGaIn is more suitable for liquid-metal UV-plasmonic applications because its melting point is well below room temperature, but the transition of pure gallium into the liquid state may require additional heating under ambient conditions. The liquid-meta EGaInl NP synthesis technique employed in this work is based on ultrasonication and is therefore simple and up-scalable. Moreover, EGaIn offers a low toxicity and has negligible vapour pressure and low viscosity,[32,33] which makes it essential for stretchable electronic and biomedical devices.[34] Therefore, in combination with strong UV plasmonic properties, excellent non-optical properties of EGaIn open up opportunities to integrate UV plasmonics with electronics, photoacoustics and optomechanics,[34] and electrochemistry[36] at the nanoscale .

**Results and discussion**

EGaIn NPs were synthesized via ultrasonication. This technique is particularly advantageous for our optical analysis because it allows producing NPs with diameters of about 25 nm to 200 nm, which have theoretically been predicted to support strong localized UV-plasmon resonances.[10] Other fabrication techniques, such as microfluidic flow focusing,[37] produce much larger EGaIn particles (diameters of ~50 μm) that cannot support any localized UV plasmon mode. Bottom-up methods that are often used to produce metal NPs are difficult to employ with gallium alloys due to its large oxidation potential.[25]

EGaIn (200 mg) was added to absolute ethanol (10 mL) and sonicated using a probe sonicator (Qsonica, Q700) for 10 minutes at 20 W in a 20 mL vial to create EGaIn particles. Larger particles precipitated for 2 hours and the supernatant was used for further experiments. The resulting particles were colloidally stable for several weeks before aggregates formed, leading to precipitation. The NP suspension was diluted (1:50 in ethanol) and drop cast onto a holey carbon TEM grid for particle size analysis and diluted 10 times in ethanol for optical analysis.

Figure 1A shows a representative TEM image of the synthesized EGaIn NPs. Based on the analysis of 264 NPs, we find that the sample contains a broad distribution of NP sizes peaking at 110 nm (Figure 1B). We find more than 20 NPs in each bin (20 nm width) in the size range from 30 nm to 170 nm. A broad NP size distribution in our sample is due to stochastic nature of the sonication process. Finally, we determine (by using both TEM and ellipsometry) that all NPs have a gallium oxide shell ($h$) of ~3 nm thickness.[21] Significantly, removing the shell would cause the NPs to coalesce and become unstable, which is highly undesirable.

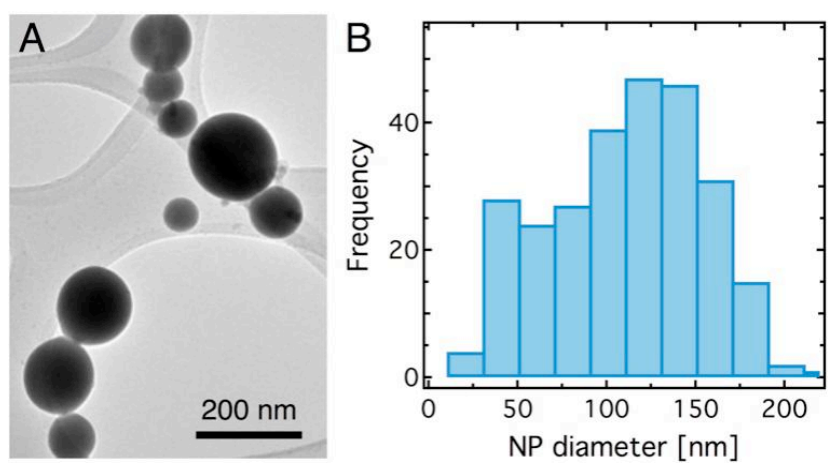

**Figure 1**. Nanoparticle size analysis. A: Representative transmission electron microscopy (TEM) image of EGaIn NPs on a holey carbon grid. B: Nanoparticle size distribution based on the TEM analysis of 264 NPs.

We performed UV-visible absorption spectroscopy experiments of the EGaIn NPs suspended in ethanol to measure the optical absorption spectra of the NPs. The NP dispersion was placed in a UV quartz cuvette for extended UV transparency. Extinction and absorption spectra (Agilent Technologies, USA, Cary 7000) were obtained using an integrating sphere for the NP dispersion and the solvent Figure 2A. Ethanol starts to absorb UV light below ~250 nm and strongly absorbs light below 205 nm, which is generally used as a lower cut-off wavelength for ethanol.[38] The ethanol contribution to the spectrum was subtracted from the NP spectra (Figure 2B).

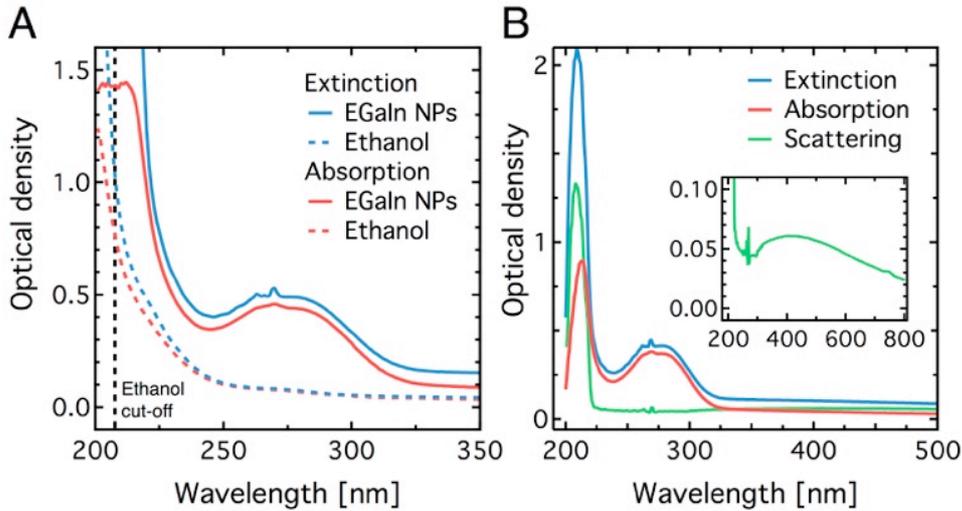

**Figure 2.** Optical spectroscopy of EGaIn NPs dispersed in ethanol. A: Raw extinction and absorption spectra of the EGaIn NPs dispersed in ethanol, with pure ethanol absorption for reference. The solvent starts to strongly absorb below 205 nm as indicated by the vertical dashed black line. B: Extinction, absorption and scattering spectra of the EGaIn NPs only. The ethanol background was subtracted and the scattering spectrum was obtained by subtracting the absorption from the extinction spectrum. The inset shows the scattering spectrum only for a wider spectral range.

The extinction and absorption spectra show two dominant peaks: one broad peak centered at ~ 275 nm and a second one at ~210 nm. Due to the close proximity of the latter peak to the cut-off wavelength of ethanol, both its absolute and relative amplitude and spectral position are potentially affected by the onset of strong ethanol absorption. By subtracting the absorption spectrum from the extinction spectrum, we find that the NPs also strongly scatter light in the region of the 210 nm peak, while the spectrum is clearly dominated by absorption above 220 nm. Another broad scattering peak is present in the visible centred at around 400 nm (see Figure 2B, inset). This confirms that, similar to liquid pure-gallium NPs,[13] EGaIn NPs dispersed in ethanol also support plasmon resonances in the visible range.

We used Mie theory[39,40] to explain the physical origin of the resonance peaks in Figure 2, and calculate the light scattering and absorption properties of single EGaIn NPs in ethanol. In general, Mie theory allows for the determination of the particles' scattering ($C_{sca}$) and absorption ($C_{abs}$) cross sections, as well as their extinction ($C_{ext} = C_{sca} + C_{abs}$) cross sections.[39,40] In the following, the light absorption is analysed in terms of a particle's absorption efficiency defined as $Q_{abs} = C_{abs}/G$,[40] where $G = \pi R^2$ is the geometrical cross-section of a sphere. Note that $Q_{abs}$ can have values greater than one, implying that the NP can absorb photons outside its geometrical cross-sectional area $G$.[39,40]

We calculate these optical properties for liquid gallium NPs with diameters $d$ of 50 nm, 100 nm and 150 nm, all of which are present in the NP ensemble (Figure 1). We take into account the experimentally measured gallium oxide shell thickness ($h = 3$ nm, Figure 3A) and assume that the NPs are embedded in a constant dielectric material of refractive index $n_{EtOH} = 1.478$, which is the refractive index of ethanol at 185 nm.[41]

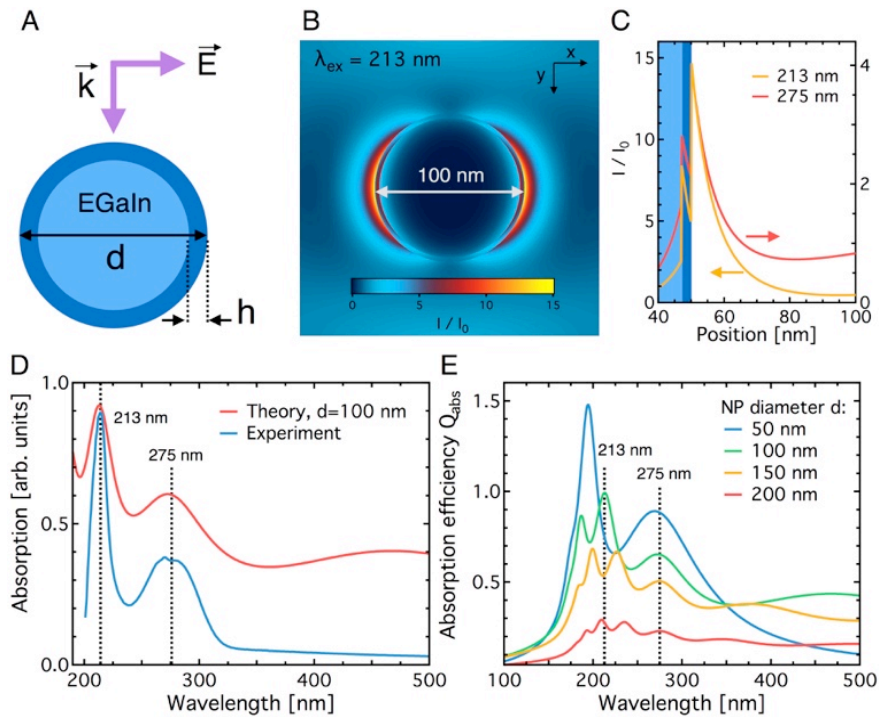

**Figure 3**. Theoretical results and comparison with experimental data. A: Geometry of the investigated EGaIn NP with diameter $d$, surrounded by a gallium oxide shell of thickness $h = 3$ nm. The orientation of the E-field polarisation and the k-vector are also shown. B: Normalized field intensity distribution $I/I_0$ ($= |\mathbf{E}|^2/|\mathbf{E_0}|^2$) in and around the NP upon 213 nm excitation. The distribution was also calculated for 275 nm excitation (not shown). C: Field intensity distribution as a function of radial distance from the NP centre along the x – axis. D: Comparison of the theoretical absorption spectrum of a 100 nm particle (red line) with the experimental absorption spectrum (blue line). E: Calculated absorption efficiency $Q_{abs}$ as a function of wavelength for 50 nm, 100 nm, 150 nm and 200 nm sized particles.

In the presence of the oxide layer, the optical properties of EGaIn and pure liquid gallium are similar.[21,42] In the UV-visible spectral region, the dielectric permittivity function of pure liquid gallium can accurately be represented by a simple Drude-like model.[13,21,42] Thus, as experimental data for the dielectric permittivity of EGaIn ($\varepsilon_{EGaIn}$) in the UV spectral region is not available, we employ the Drude model to extrapolate experimental values of $\varepsilon_{EGaIn}$ obtained for the visible-to-NIR (400 nm to 1000 nm) spectral region.[21]

Figure 3B shows the normalized field intensity distribution $I/I_0$ ($=|\mathbf{E}|^2/|\mathbf{E_0}|^2$) at a wavelength of 213 nm, where I and $I_0$ denote the calculated and incident field, respectively. Here, a close to 15-fold field intensity enhancement is reached at the surface of the NP, while for 275 nm excitation this enhancement is only 4-fold (Figure 3C). The radial field intensity cross-section in Figure 3C also shows that the field intensity is more strongly confined to the NP surface at 213 nm compared to 275 nm. From studies of conceptually similar elongated plasmonic NPs,[43,44] it is known that a stronger field confinement in the near-field zone of the NP leads to a lower radiative broadening of the resonance peak detected in the far-field zone. Indeed, in good agreement with our experimental observations, the theoretical resonance peak at 213 nm is significantly narrower than the peak at 275 nm (Figure 3D). Figure 3C also reveals that the dielectric properties of the thin oxide layer result in an additional obstacle for the enhanced field to penetrate the metal surface of the NP, which is a well-known effect demonstrated in systems with a technologically important artificial thin dielectric layer.[45,46]

For a 100 nm particle, the experimentally observed and calculated absorption peak positions are in excellent quantitative agreement and located at 213 nm and 275 nm (Figure 3D). The peak at 213 nm shows a narrow spectral width of 9 nm (experiment) and 11 nm (theory). The broader peak at 275 nm has a linewidth of 92 nm (experiment) and 26 nm (theory). The number and intensity of resonance peaks in the spectral region between 150 nm and 250 nm varies for the investigated NP sizes in the size range of 50 nm - 200 nm (Figure 3E). While the calculated spectrum for a 50 nm particle shows only one pronounced peak below 200 nm (experimentally not accessible here), the spectrum of a 200 nm particle exhibits three far less pronounced peaks. Importantly, all particle sizes show a significant peak at 275 nm in agreement with experiments.

## Conclusions

We have investigated the UV plasmonic properties of eutectic gallium-indium (EGaIn) liquid-metal NPs suspended in ethanol and calculated the particles' optical properties using Mie theory. We experimentally demonstrate that the NPs exhibit UV plasmonic resonances. In agreement with our theoretical model for a 100 nm particle, we show that the two main resonances are located at 213 nm and 275 nm. These UV plasmonic properties of liquid-metal NPs are of significant interest due to their potential applications in biomedical imaging, sensing and medicine. The results presented in this work also further contribute to the concept of reconfigurable liquid-metal plasmonics,[35] where plasmon resonances of liquid-metal NPs can be tuned by deforming the shape of the NP using, for example, ultrasound. In this case, the NPs must be in solution to enable an efficient propagation of ultrasound from the source to the NPs. Importantly, although the plasmonic properties of liquid-metal NPs made of pure gallium are similar,[25] the lower melting point of EGaIn (~15.5 °C) compared to pure Ga (~30 °C) for example make it more suitable for applications at room temperature. Finally, an increasingly important role of liquid metal EGaIn NPs in stretchable electronic devices and mechanics[34] opens up opportunities to further integrate plasmonics with electronics, optomechanics and electrochemistry[36] at the nanoscale .

## Acknowledgements

This work was supported by Australian Research Council (ARC) through its Centre of Excellence for Nanoscale BioPhotonics (CE140100003), LIEF program (LE160100051) and Future Fellowship (FT1600357). This research was undertaken on the NCI National Facility in Canberra, Australia, which is supported by the Australian Commonwealth Government. P.R. acknowledges funding through the RMIT Vice Chancellor's Research Fellowship. The authors acknowledge the support of the Ian Potter Foundation in establishing the Ian Potter NanoBioSensing Facility at RMIT University, where the UV-Vis absorption spectroscopy was carried out. Y.L. and M.D.D. acknowledge funding from the National Science Foundation through the Research Triangle MRSEC (DMR-1121107). The authors acknowledge use of the Analytical Instrumentation Facility (AIF) at North Carolina State University, which is supported by the State of North Carolina and the National Science Foundation (award number ECCS-1542015).

**Author contribution statement**

I.M. performed all calculations and simulations. P.R. performed all optical experiments and prepared all graphs. Y.L. synthesized the investigated particles and performed electron microscopy analysis. P.R. and I.M. designed experiments and drafted the manuscript. All authors contributed to writing and revising the manuscript and contributed to analysing and interpreting the experimental results.

**Conflicts of interest**

All authors declare no conflict of interest.